\documentclass[twocolumn,floatfix,superscriptaddress,amsmath,showpacs,aps,prb]{revtex4}

\usepackage{t1enc}
\usepackage[final]{graphicx}
\usepackage{epsfig}
\usepackage{bm}
\usepackage[normalem]{ulem}
\usepackage[usenames]{xcolor}
\usepackage{times}
\usepackage{verbatim}
\usepackage{amssymb}

\begin{document}

\title{On the mechanism behind the inverse melting in systems with competing interactions}

\author{Alejandro Mendoza-Coto}
\email{alejandro.mendoza@ufsc.br}
\affiliation{Departamento de F\'\i sica, Universidade Federal de Santa Catarina, 88040-900, Florian\'opolis, SC, Brazil}
\author{Lucas Nicolao}
\email{lucas.Nicolao@ufsc.br}
\affiliation{Departamento de F\'\i sica, Universidade Federal de Santa Catarina, 88040-900, Florian\'opolis, SC, Brazil}
\author{Rogelio D\'iaz-M\'endez}
\affiliation{Department of Theoretical Physics, KTH Royal Institute of Technology, SE-106 91 Stockholm, Sweden}






\date{\today}

\begin{abstract}

Inverse melting are those in which the more symmetric phase (e.g. paramagnetic or fluid phase) is 
established at lower temperatures, while the less symmetric phase (e.g. modulated 
or crystalline phase) appears at higher temperatures. This exotic scenario has been observed in the 
reentrant phase diagrams of systems with competing interactions. In such systems, the competition between a short 
range attractive interaction and a nonlocal repulsive interaction promote the appearance
of modulated phases.
Here we present a fundamental comprehension of the microscopic mechanisms leading to the emergence of such inverse 
transitions by considering a thorough mean-field analysis of a variety of minimal models with different competing interactions. 
Through analytical and numerical tools we identify the specific connections between the characteristic energy of the homogeneous 
and modulated phases and the observed reentrant behaviors. In particular, we find that reentrance is appreciable when the 
characteristic energy cost of the homogeneous and modulated phases are comparable to each other, and for systems in which the 
local order parameter is limited.
In the asymptotic limit of high energy cost of the homogeneous phase we obtain analytically that the degree 
of reentrance of the phase diagram decreases exponentially with the ratio of the characteristic energy cost of homogeneous and 
modulated phases.
We are also able to establish theoretical (upper and lower) bounds for the degree of the reentrance, according to the 
nature of the competing interactions.
Finally, we confront our mean-field results with Langevin simulations of an effective coarse grained model, confirming the main 
results regarding the degree of the reentrance in the phase diagram.
These results shed new light on the many systems undergoing inverse melting transitions, from magnets to colloids and vortex matter,  
by qualitatively improving the understanding of the interplay of entropy and energy around the inverse melting points.   
\end{abstract}






\maketitle

\section{Introduction}
\label{Sec:Introduction}

The inverse melting (IM) is generally understood in terms of a larger entropic contribution arising when the systems 
go into a less-symmetric phase by increasing the temperature \cite{Greer2003}.
This is in clear contradiction with the commonly extended association between symmetry and order, i.e. less symmetry implies 
more order and vice-versa.
It is therefore highly counter-intuitive the many experimental findings verifying IM transitions in which, by 
augmenting the temperature, liquids become solids and homogeneous magnetization becomes modulated patterns \cite{Dudowicz2009,Frenkel2012,lang2015,Sa2016}.
There is no fundamental contradiction whatsoever, since the actual measure of order is given precisely through the entropy and not 
the symmetry. 
In this way, such anomalous transition can happen, for instance, when some degrees of freedom are frozen in the
liquid phase; under that circumstance the IM can lead to a gain of entropy by exciting these degrees of freedom in a solid 
phase \cite{Debenedetti03}.
Several models have been used to reproduce such reentrant behaviors in particle systems \cite{almudallal2014}, spin systems 
\cite{Schupper04,Crisanti2005,Tarzia16} and coarse grained approaches \cite{Buceta01}. Even though in some of these systems 
the connection of the reentrant behavior with the specific features of the interactions and microscopic details has been clarified,
there is a large class of systems with competing interactions in which such connection is yet to be established. Here we address 
the case of systems where isotropic competing interactions gives rise to modulated phases in two dimensions.

Modulated phases typically appear in systems where a short-range
attractive interaction competes with a non-local repulsive
interaction. In two dimensions is commonly observed stripes and
bubbles, or clusters, configurations with a modulation length
depending on the relative strength of the interactions
\cite{SeAn1995}. Systems with this kind of phenomenology are present
from soft condensed matter to magnetic systems. For instance, in
binary (AB) polymer mixture in which the polymer chains are connected
by a covalent bond at the chain ends, microphase separation takes
place. In this case, A-rich and B-rich domains can arrange them selves
in lamelar structures of size of the order of the sum of each polymer
chain \cite{Bates1999,Harrison2000,Ohta2009}. In charged colloidal systems phase
separation is possible due to the competition between the short range
attractive Van der Waals interaction and the non-local repulsive
screened Coulomb interaction
\cite{Sciortino2004,Stradner2004,Zaccarelli2007}. 
On the other hand,
ferromagnetic dipolar thin films present modulated structures due to
the competition between the ``attractive'' exchange interaction and the
``repulsive'' dipolar interaction \cite{Cannas2006,Back2015}.

In some of these systems the IM transitions have been predicted
theoretically \cite{Andelman1987} or even observed experimentally
\cite{PoVaPe2003}.
In the magnetic case, a IM transition has been recently observed in
experiments on ultrathin ferromagnetic films with perpendicular
anisotropy of Fe/Cu(001) \cite{Sa2010Fe, Sa2010No, Sa2016}, which have
shown that a perpendicular magnetic field versus temperature phase
diagram displays strong reentrant features - a behavior somewhat
predicted theoretically for this particular magnetic systems
\cite{abanov95}. In an attempt to explain this experimental
observations, a scaling theory was developed to relate the presence of
a certain family of microscopic interactions with the existence of
reentrant behavior in the phase diagram \cite{Po2010}. More recently,
the same reentrant behavior have been found within mean-field
approximations for systems with dipolar repulsive interactions in the
context of Landau-Ginzburg models \cite{cannas11}, lattice models
\cite{velasque14} and others \cite{mc2016}.  Many of these efforts
focus on the relation between the IM behavior with an entropy gain
from domain wall degrees of freedom of the modulated patterns.  In
spite of these advances, little progress have been made relating the
nature of the microscopic interactions and the underlying mechanisms
of the IM transitions in the phase diagram.

In the present work we are able to explore this relation by generalizing a previously developed minimization-variational 
technique \cite{mc2016} to consider generic isotropic interactions and different approximations schemes. 
Different competing interactions, encoded in $A(r)$, are explored by altering the 
functionality of the fluctuation spectrum $\hat{A}(k)$, 
which is the Fourier transform of the interactions of the system. This is the key quantity 
that should contain the basic ingredients for the reentrant behavior to be present. 
Indeed, we found that, in order to observe an IM transition, the relative energy cost of the homogeneous 
phase must comparable to the 
characteristic energy cost of the modulated phase, in other words $\hat{A}(0)\simeq \hat{A}(k_0)$, 
where $\hat{A}(k_0)$ stands for the minimum of the fluctuation spectrum.

Furthermore, we found that another key ingredient for the IM
transition to take place is that the local entropy has to be a steep
enough function of the order parameter close to its saturation value.
Physically, the steepness of the functional form of the local entropy
establishes a constraint in the order parameter. In other words, we
need that the order parameter has to be limited, not necessarily by a
hard constraint. This implies for instance that in a magnetic system
we must have a saturation value for the local magnetic moment and in
the case of a fluid, particles must have some kind of hard core or
another type of strong enough short-range repulsive interactions.

In order to understand the relation between the reentrance behavior in the phase diagram and the 
underlying microscopic model, we have considered several specific fluctuation spectra, corresponding to 
different types of microscopic interactions relevant for several physical systems 
(see table \ref{table} in Section~\ref{meh}). 
We found that the extension of the IM 
in the phase diagram is proportional to the quantity $\hat{A}(0)/\hat{A}(k_0)$. 
In the limit where this ratio is large, we obtained an
analytical expression characterizing the extension of the reentrance for a given 
phase diagram, independently of the details of the microscopic model.
We also found that the extension of the reentrance has a theoretical upper and lower bound, depending only on the 
the general features of the fluctuation spectrum $\hat{A}(k)$ of the specific model.
Finally, the overall key results, i.e. the shape of the phase diagrams as a function of the form of the interactions, are 
additionally contrasted with results of Langevin simulations for the ferromagnetic dipolar model. The latter, confirming the validity of the
theoretical approaches, are the first computer simulations to verify the IM transition in coarse-grained models.

The paper is organized as follows, in section 1 we present the generic microscopic model to be studied and its coarse grained version used 
along this work; we also present the variational methods employed to study the equilibrium state of the system. 
In section 2 we used some specific fluctuation spectra to study under which condition reentrance is observed and discuss how the reentrance 
of a given model is limited superior and inferiorly. Section 3 is devoted to a discussion in depth of how inverse melting actually takes place 
and how the interplay energy versus entropy is responsible for it. In the fourth section we present the result of Langevin numerical simulation for
a specific model validating the main conclusions obtained through analytical calculations. Finally in the last section of the paper we present 
the conclusions of our work.  

\section{Theoretical Approach}

Let us start by considering an Ising-like spin system $\{s_i\}$ with generic non-local interactions $A_{ij}$ in a two dimensional square lattice. 
This Hamiltonian can be written in the form:
\begin{equation}
 \mathcal{H}=\frac{1}{2}\sum_{i,j}A_{ij}s_is_j-\sum_i h s_i,
\label{eq1} 
\end{equation}
where $h$ is an applied external field.
Within this general expression, we will focus on those interactions that are able to generate spatially modulated 
patterns in the equilibrium regime of the system. 

Typically, these spatial textures are in form of stripes or bubbles patterns, which can extend over a large number of lattice sites.  
In dipolar frustrated ferromagnetic materials, for instance, the typical stripe width is about thousand times the 
lattice spacing \cite{saratz2016}. 
On the other hand, theoretical and experimental studies on systems whose modulation is of the order of 
the minimum distances between their constituents have encounter no evidence of IM \cite{dm10}. 
This scenario  fully justify the use of coarse-grained approaches to study the IM transition. 

After a coarse-graining extension of Eq.~(\ref{eq1}), the new effective 2D model is now described by a scalar-field local-order parameter $\phi(\vec{x})$, 
with effective Hamiltonian written in the form \cite{Me2012,mc2016}:
\begin{align}
\nonumber
H[\phi] &= \frac{1}{2} \iint d^2x d^2x' \phi({\vec x})\phi({\vec x'})A(\left|{\vec x}-{\vec x}'\right|) \\
&+ \frac{1}{\beta} \int d^2x\;  S(\phi({\vec x}))
- \int d^2x\; h \phi({\vec x}), 
\label{eq2}
\end{align} 
where the generic interaction $A({\vec x},{\vec x}')=A(\left|{\vec x}-{\vec x}'\right|)\equiv A(r)$ is considered isotropic, $\beta$ stands for $(KT)^{-1}$ and   
\begin{align}
S(x)=\frac{(1+x)}{2}\ln\left(\frac{1+x}{2}\right)+\frac{(1-x)}{2}\ln\left(\frac{1-x}{2}\right)
\label{eqS}
\end{align}
represent a local potential for the continuous order parameter, which can be seen as the microscopic entropic contribution to the coarse grained model.
In fact, this form of the Hamiltonian is a local density approximation on the mean-field free energy
of the model of Eq.~(\ref{eq1}).

In the following, we study the equilibrium phase diagrams of the Hamiltonian of Eq.~(\ref{eq2}).
This is done by looking at the stationary states of the overdamped Langevin equation of the system in terms of
the local order parameter $\phi(x)$.
Considering Eq.~(\ref{eq2}), a natural choice for the dynamical equation reads
\begin{align}
\nonumber
\frac{\partial \phi(\vec{x},t)}{\partial t}&=-\frac{\delta H[\phi]}{\delta \phi(x)}+\eta(\vec{x},t)\\
\nonumber
&=- \int d^2x' A(\left|{\vec x}-{\vec x}'\right|) \phi({\vec x'},t)\\
&- \frac{1}{\beta}\ \mathrm{arctanh}(\phi(\vec{x},t)) +h+\eta(x,t),
\label{eq3}
\end{align}
where $\eta(\vec{x},t)$ represent the usual zero-mean white noise, with 
$\left<\eta(\vec{x},t)\eta(\vec{x}',t')\right>=2\beta^{-1}\delta(t-t')\delta(\vec{x}-\vec{x}')$. 

Within the mean field approximation, the stationary state of Eq.~(\ref{eq3}) implies that  
$\partial_t\langle\phi(\vec{x},t)\rangle$ must be zero. 
Angular brackets indicate the thermal average of the corresponding quantity. 
Taking the thermal average over both sides of Eq.~(\ref{eq3}) we obtain
\begin{equation}
\frac{1}{\beta}\ \langle\mathrm{arctanh}(\phi(\vec{x},t))\rangle =
- \int d^2x' A(\left|{\vec x}-{\vec x}'\right|) \langle\phi({\vec x'},t)\rangle
+h.
 \label{eq4}
\end{equation}
At this point a relation between $\langle\mathrm{arctanh}(\phi(\vec{x},t))\rangle$ and 
$\langle\phi({\vec x},t)\rangle$ is necessary, and 
different approximations could be used \cite{Me2012}.
In the present work, we begin by using the standard mean-field approach, neglecting all fluctuations in the local 
order parameter: 
$\langle\mathrm{arctanh}(\phi(\vec{x},t))\rangle\cong\mathrm{arctanh}(\langle\phi(\vec{x})\rangle)$. 
In this case we can recast Eq.~(\ref{eq4}) as:
\begin{equation}
\frac{\delta H[\langle\phi\rangle]}{\delta \langle\phi(x)\rangle}=0.
\label{eq5}
\end{equation}
This means that, in the crude mean field approach, the equilibrium state of the system corresponds 
to configurations minimizing the effective
coarse grained Hamiltonian $H[\phi]$, i.e. the mean field free energy of the original microscopic 
model of Eq.~(\ref{eq1}). 

In order to test the robustness of the phase diagrams obtained within this approximation, we additionally 
calculate an improved version of this mean-field approach.
This is accomplished by including the equilibrium non-linear dynamics of the local order 
parameter in Eq.~(\ref{eq3}). 
Let us then consider the exact Langevin equation of motion for a single site of the system. 
According to Eq.~(\ref{eq3}) we can write: 

\begin{equation}
 \frac{\partial \phi(\vec{x},t)}{\partial t}=h_{{l}}(\vec{x},t)- 
 \frac{1}{\beta}\ \mathrm{arctanh}(\phi(\vec{x},t)) +\eta(\vec{x},t),
 \label{eq6}
\end{equation}
where $h_{{l}}(\vec{x},t)$ represents the local field in the equilibrium state. 
An improved mean field approximation can be devised by neglecting fluctuations in the local field. 
In this approximation the local field is simply taken as a constant along time, equal to the mean 
local field $h_{\mathrm{eff}}(\vec{x})\equiv\left<h_l(\vec{x},t)\right>$. Once that we have decoupled 
the multiple single site equations, we can solve exactly the stochastic equation for generic single 
site. The single site Langevin equation resulting from Eq.~(\ref{eq6}) has a stationary equilibrium 
state satisfying the Boltzmann probability distribution, thus
\begin{equation}
 P(\phi)=\frac{1}{Z}\exp[-\beta(-h_{\mathrm{eff}}\phi+\frac{1}{\beta}S(\phi))],
 \label{eq7}
\end{equation}
where $Z$ is a normalization constant. 

Once the form of the probability distribution $P(\phi)$ is known, it is possible to calculate
$\langle \mathrm{arctanh}[\phi]\rangle$ and $\langle\phi\rangle$ as a function of the parameter $\beta h_{\mathrm{eff}}$. 
This allow to parametrically obtain the function $\langle \mathrm{arctanh}\rangle(\langle\phi\rangle)$.
In turn, this relation ($\langle \mathrm{arctanh}\rangle(\langle\phi\rangle)$) can be used in Eq.~(\ref{eq4}) to obtain the
improved mean-field description of the model. 
Interestingly, the final equation can still be written in the same form:
\begin{equation}
\frac{\delta H_1[\langle\phi\rangle]}{\delta \langle\phi(x)\rangle}=0,
\label{eq8}
\end{equation}
using the improved version of the Hamiltonian:
\begin{align}
\nonumber
H_1[\phi] &= \frac{1}{2} \iint d^2x d^2x' \phi({\vec x})\phi({\vec x'})A(\left|{\vec x}-{\vec x}'\right|) \\
&+ \frac{1}{\beta} \int d^2x\;  S_1(\phi({\vec x}))
- \int d^2x\; h \phi({\vec x}), \label{eq9} 
\end{align}
with the effective entropic contribution:  
\begin{align}
S_1(\phi)=-\int_\phi^1\ dx\ \langle \mathrm{arctanh}\rangle(x).
\label{eqS1}
\end{align}

\begin{figure}[th!]
  \centering
\includegraphics[width=0.9\columnwidth]{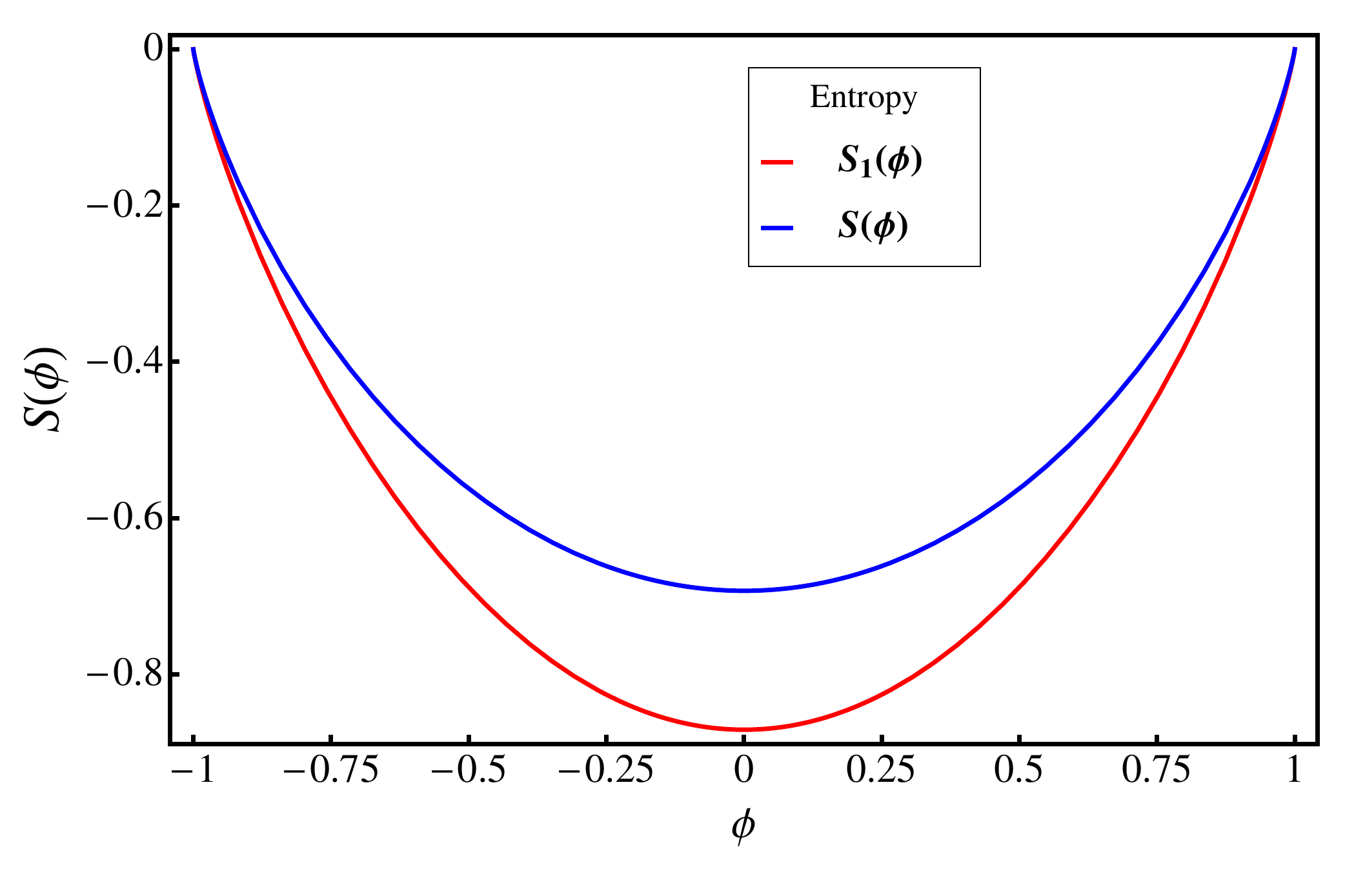}
\caption{Free energy functionals $S(\phi)$ and $S_1(\phi)$, as defined in Eq.~(\ref{eqS}) and Eq.~(\ref{eqS1}).}
\label{fef}
\end{figure}

Again, this means that the equilibrium state can be obtained by minimizing certain free energy functional.  
For comparison, in Fig~\ref{fef} we show the corresponding entropic functions in both approximations. 
As can be seen, once we include the nonlinear local dynamics there is an entropy gain due to the soft nature 
of the order parameter - it is worth noting that local entropy is defined as $-S(\phi)$ in Eqs.~(\ref{eq2},\ref{eq9}).

Once these free energy functionals of Eq.~(\ref{eqS}) and Eq.~(\ref{eqS1}) have been defined, we can proceed 
with the minimization process to study the equilibrium states of the system.

\section{Minimization of the effective Hamiltonians}
\label{meh}

\mbox{}
\indent
The minimization technique we perform in this work is analogous to that used in Ref.~\cite{mc2016}.
We are interested in solutions that minimize the effective free energy. 
We consider three types of solutions: $i)$ {\it stripes}, where $\phi(\vec{x})$ is given by a one 
dimensional modulation; $ii)$ {\it bubbles}, where a two dimensional modulation occurs in the form of 
a triangular array of opposite values of $\phi$ in relation to the background value; and $iii)$ {\it uniform}, 
where $\phi$ takes a constant homogeneous value. 
The general solution for such configurations can be written as:
\begin{align}
 \phi(\vec{x})=\sum_{i=0}^n c_i\  \mathrm{cos}(\vec{k}_i\cdot\vec{x}),
 \label{sol}
\end{align}
where the set of wave vectors $\vec{k}_i$ are conveniently taken in order to reproduce the different solutions.
Here $n$ is the number of wave vectors considered in the solution, once the number of modes along the principal 
direction have been fixed. In the case of the stripes solution we have considered $n=15$ and this corresponds 
to the number of modes in the principal direction. In the case of the bubbles solution, where we have three 
principal directions $(1,0),(-\frac{1}{2},\frac{\sqrt{3}}{2}),(-\frac{1}{2},-\frac{\sqrt{3}}{2})$, we have $n=360$ 
wave vectors when considering $15$ modes in each principal direction. 
The wave vectors are defined as $\vec{k}_i=k_\mathrm{eq}\vec{a}_i$, where $k_\mathrm{eq}=2\pi/\lambda_\mathrm{eq}$ 
corresponds to the modulation length at equilibrium $\lambda_\mathrm{eq}$, and the set of vectors $\vec{a}_i$ 
are chosen as a regular 1D (stripes) or 2D triangular (bubbles) lattice, of spacing equal to 1.

\begin{table}[h] 
\centering 
\begin{tabular}{|c| c|}
\hline 
Name& Expression\\[0.4ex] 
\hline 
quadratic model& $\hat{A}(k)=a(k-1)^2-1$ \\
\hline 
quartic model& {$\hat{A}(k)=a(k^2-1)^2-1$} \\ 
\hline 
screened Coulomb model& $\hat{A}(k)=-2(2+a^2)+k^2+\frac{2(1+a^2)^{3/2}}{\sqrt{k^2+a^2}}$ \\    
\hline
{single-mode approximation}& { 
$ 
\hat{A}(k) =
  \begin{cases}
  A_0     & \text{if $k=0$}   \\
  -1 & \text{if $k=1$} \\
  + \infty       & \text{elsewhere}
  \end{cases}
$
}  
\\
\hline
\end{tabular}
\ \\ \ \\ \ \\
\caption[]{\normalfont Analytical forms of several fluctuation spectra $\hat{A}(k)$ leading to modulated configurations, 
as are treated in this work. In all cases the minimum has been set to $\hat{A}(k_0=1)=-1$, so that the remaining 
free parameter $(a)$ determines the value of $\hat{A}(0)$.
}
\label{table}
\end{table}

Assuming a local order parameter in the form of Eq.~(\ref{sol}), the functional Hamiltonians 
in Eq.~(\ref{eq2}) and Eq.~(\ref{eq9}) becomes then functions of $k_\mathrm{eq}$ and the set of 
amplitudes $\{c_i\}$, for each kind of solution. Such functions are minimized 
to find the best parameters corresponding to the three type of solutions 
for a given point in the $H$-$T$ space. The solution with minimal free energy is used to construct the 
$H$-$T$ phase diagram.

This process is carried out for some representative interactions of this class of pattern forming systems 

$A(r)=\int\frac{d^2k}{(2\pi)^2}\hat{A}(k) e^{i\vec{k}\cdot\vec{r}} $, where the function $\hat{A}(k)$  is the 
so called {\it fluctuation spectrum}. 
In order to develop modulated structures, $\hat{A}(k)$ must have a negative minimum at some nontrivial wave 
vector $k_0\neq0$. In Table~\ref{table} we list the analytic forms of the fluctuation spectra considered in this
work. As discussed above, these interactions correspond to several models physically relevant for condensed matter.
The quadratic fluctuation spectrum has been extensively used to study different aspects of the Ising ferromagnet with dipolar interactions \cite{Jagla2004,MeStNi2015}. 
On the other hand, the quartic spectrum is the continuous limit of a system with competing first-neighbors attractive and 
second-neighbors repulsive interaction \cite{Jin2012}, this kind of fluctuation spectrum have been used in effective models 
of diblock copolymers in the weak segregation limit \cite{Christensen1998,Leibler1980}. 
We also consider a model where a short range attraction competes with a repulsive potential of Yukawa type, which has been used 
to model cluster forming colloidal systems \cite{Tarzia2006}.
In addition, we analyze the single mode case, which is a minimal prototype fluctuation spectrum of those 
considered here. For this interaction the restriction in the number of modes is included in the model by setting 
to infinity the energy cost of all but the zero and the principal modes. Though unrealistic, this model becomes 
a benchmark for our theoretical calculations.

\begin{figure}[th!]
    \centering
\includegraphics[width=0.46\columnwidth]{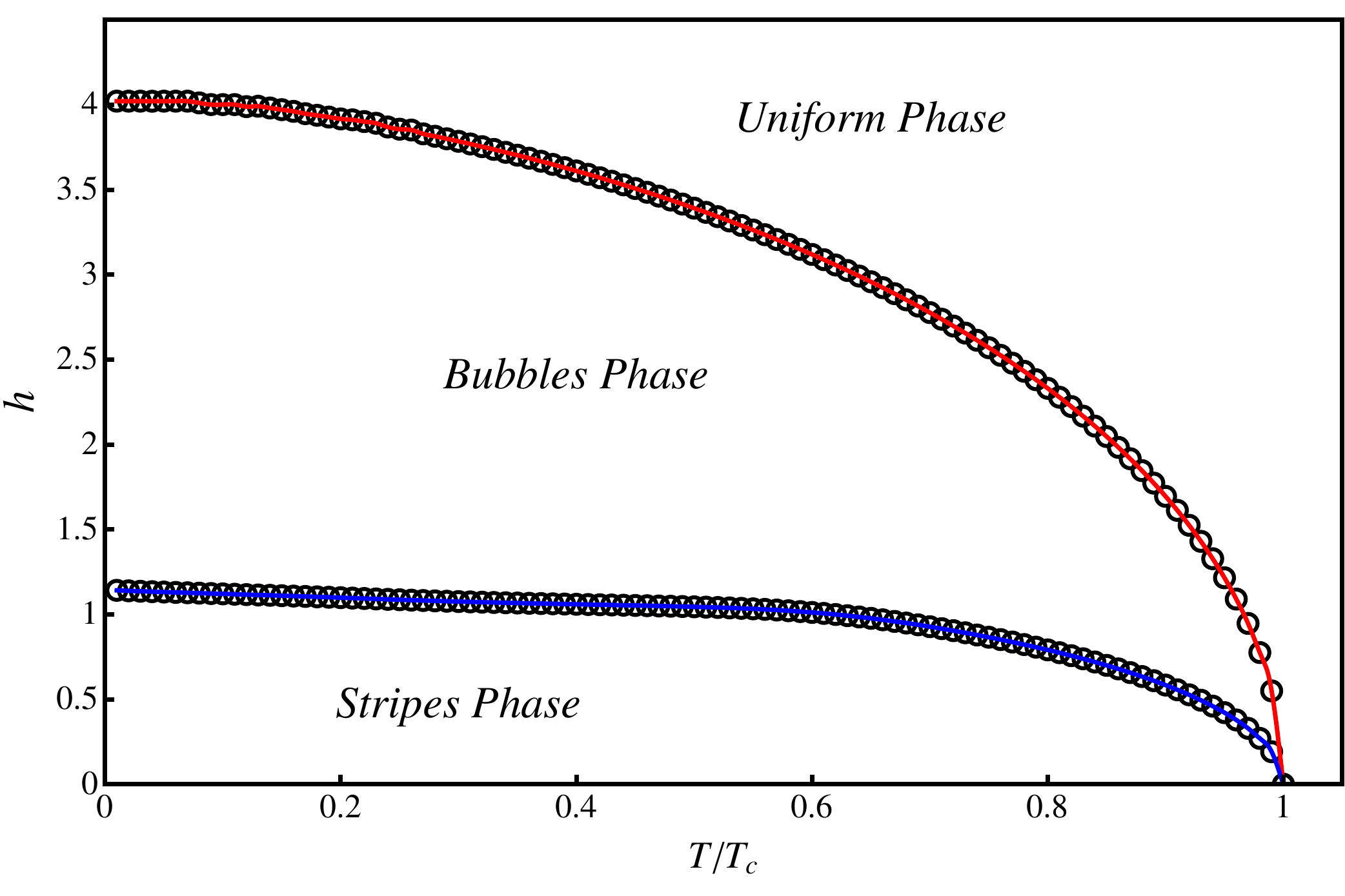}
\includegraphics[width=0.46\columnwidth]{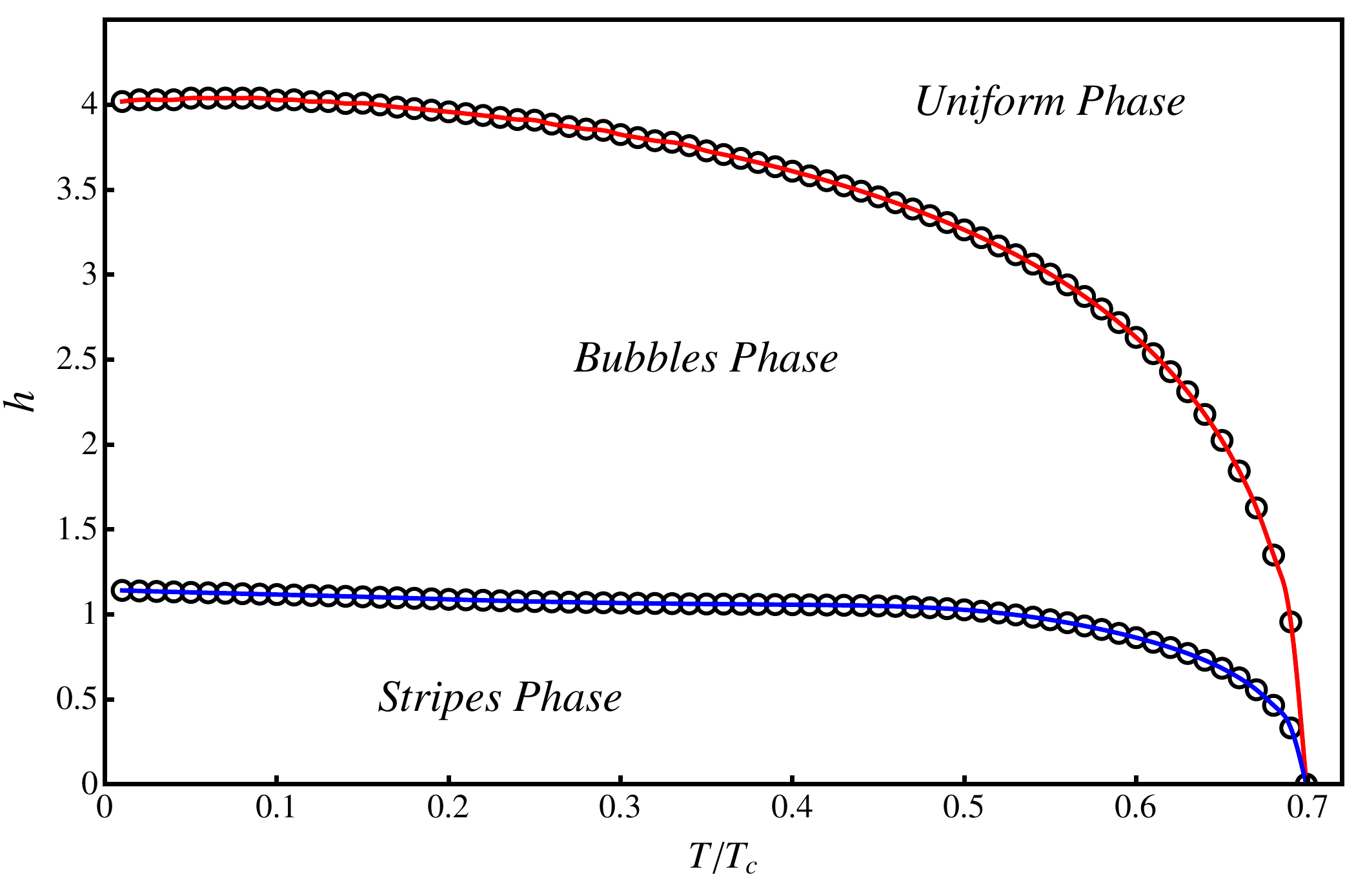}\\
\includegraphics[width=0.46\columnwidth]{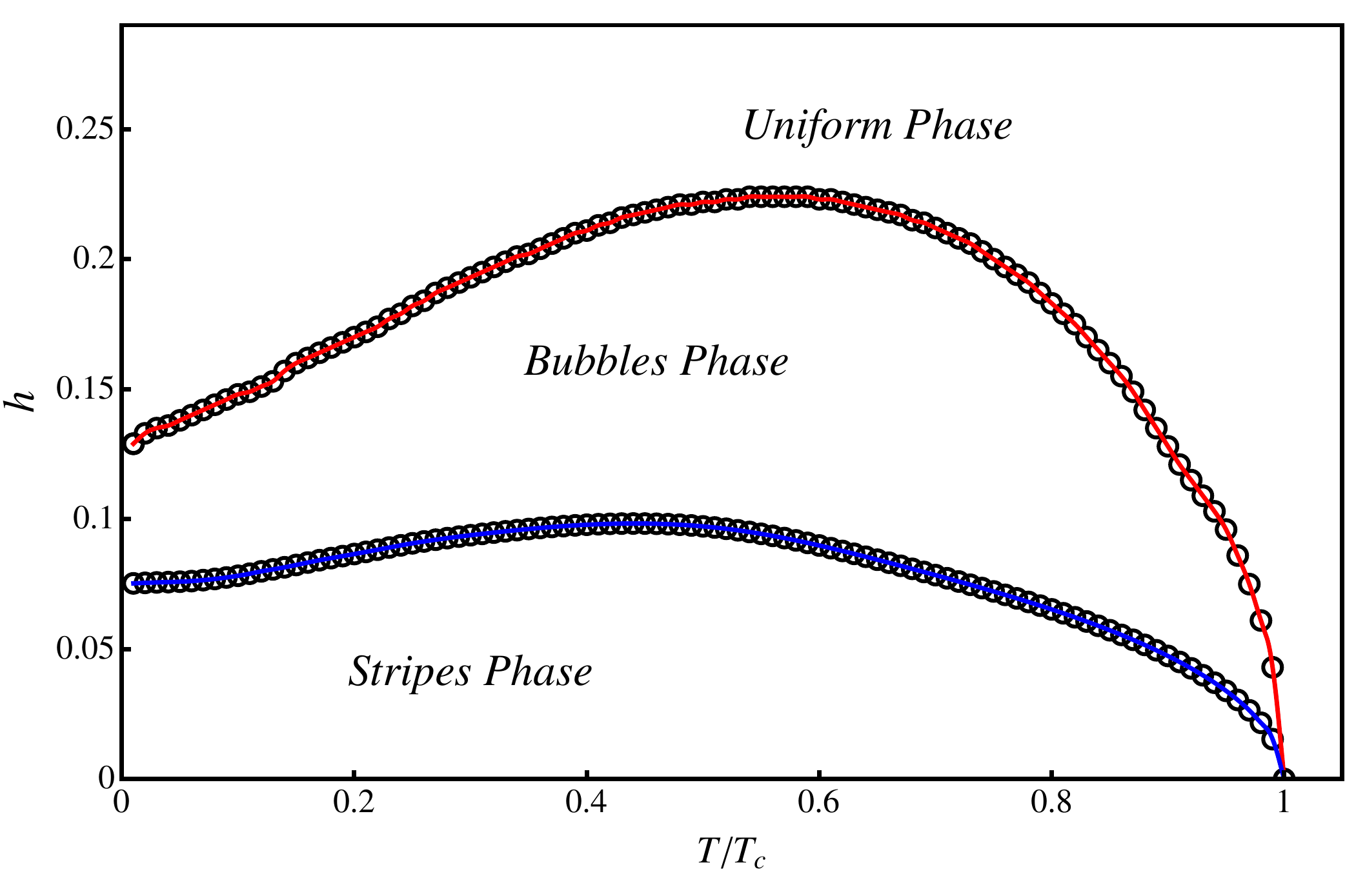}
\includegraphics[width=0.46\columnwidth]{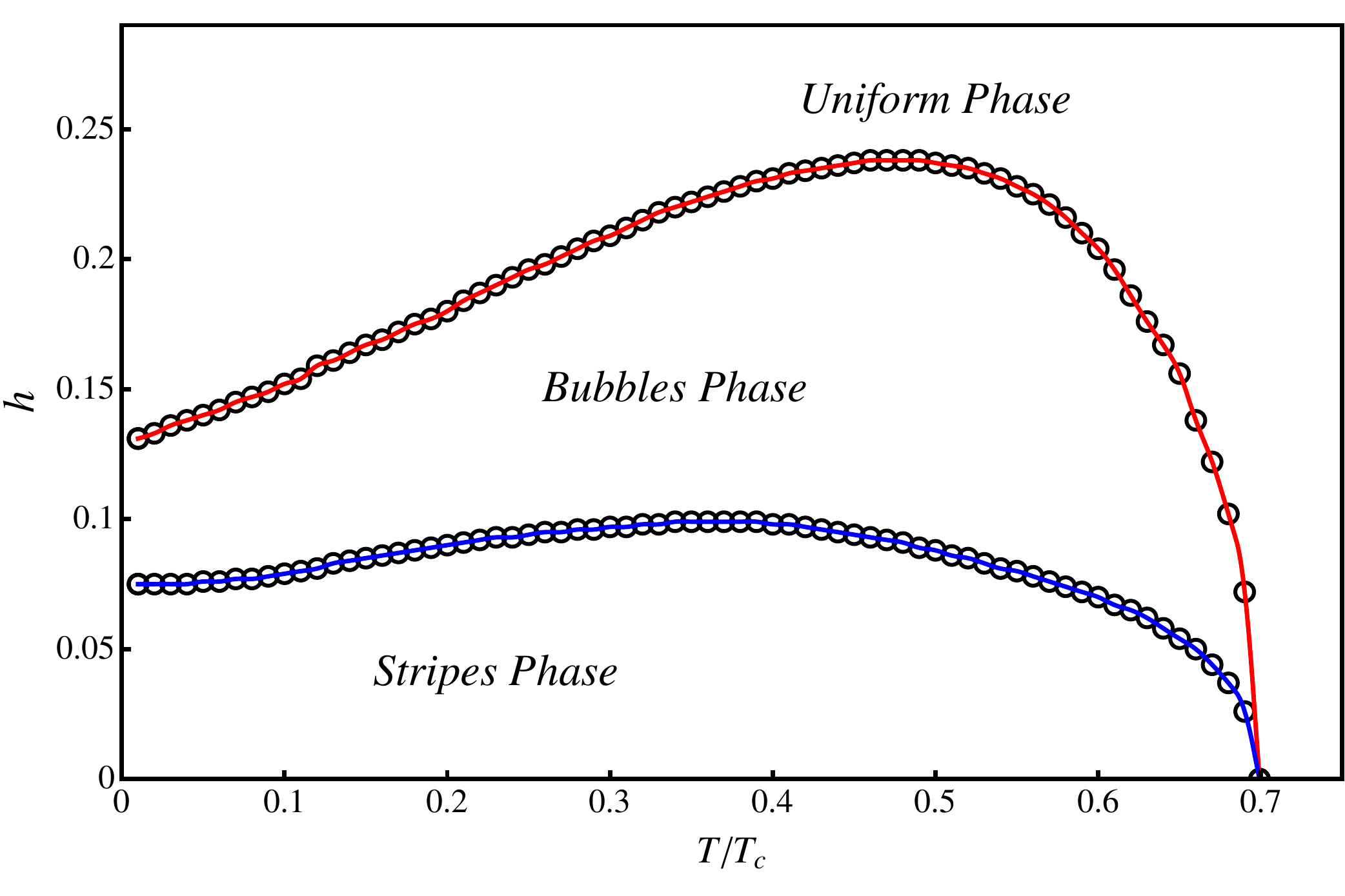}\\
\includegraphics[width=0.46\columnwidth]{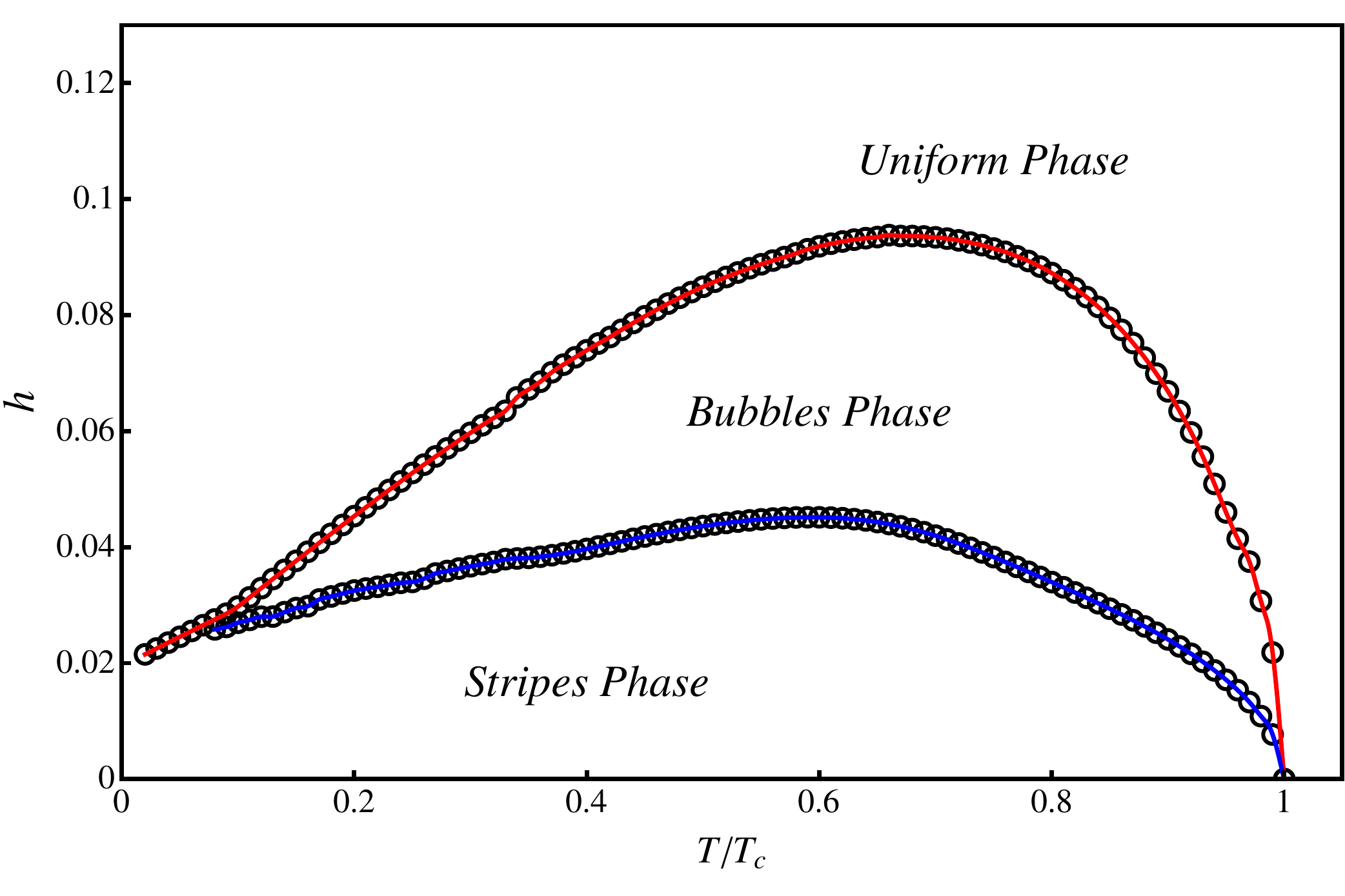}
\includegraphics[width=0.46\columnwidth]{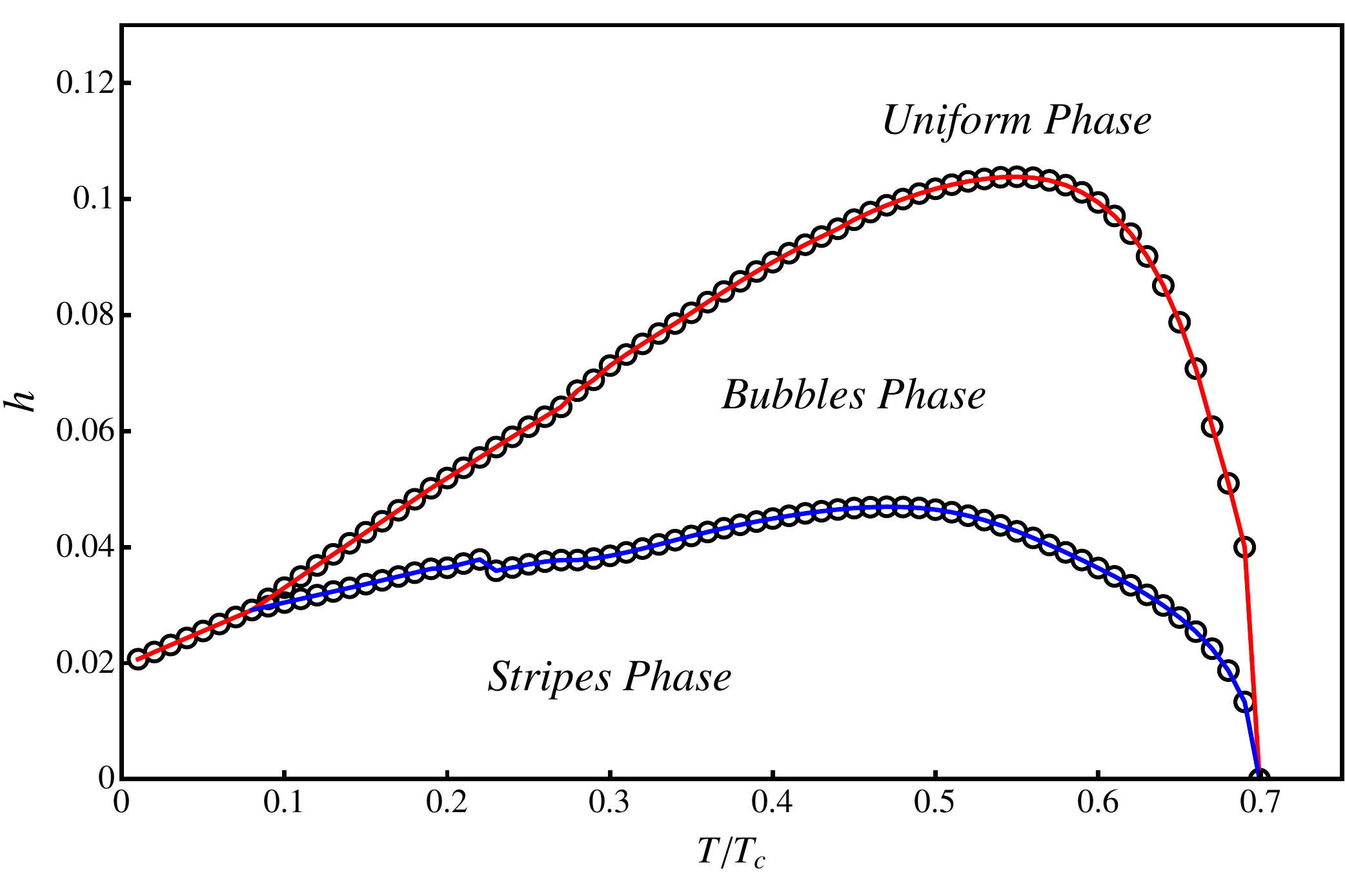}\\
\caption{
Phase diagrams corresponding to a system with fluctuation spectrum of the quadratic type, setting the number of modes 
in the principal directions to $n_\mathrm{max}=15$.
Rows corresponds to single values of the fraction $\hat{A}(0)/\vert\hat{A}(k_0)\vert$; with $\hat{A}(0)/\vert\hat{A}(k_0)\vert=4$ for panels 
A and B, $\hat{A}(0)/\vert\hat{A}(k_0)\vert=-0.6$ for panels C and D, and $\hat{A}(0)/\vert\hat{A}(k_0)\vert=-0.8$ for panels E and F.
Columns correspond to different approaches: diagrams in panels A, C and E calculated via mean field, and diagrams 
in panels B, D and F calculated with the improved mean-field minimization.
}
\label{fig1}
\end{figure}

Results for the phase diagrams of the quadratic model are shown in
Fig.~\ref{fig1}.  Firstly, it is remarkable that improved mean field
(right column of Fig.~\ref{fig1}: panels B, D and F) and strict mean
field approximations (left column of Fig.~\ref{fig1}: panels A, C and
E) yield qualitatively similar phase diagrams. It is worth noting that
the critical temperatures in the improved mean field case are lower
than the mean field ones - as expected, since in the former case local
fluctuations are included. Additionally, the values of the critical
fields as a function of the reduced temperature are very close.
Considering that the strict mean field approach has been used more
often in previous works \cite{Me2012,mc2016}, in what follows the
overall presented analysis have been obtained within this
approximation.

Interestingly, it can also be observed from Fig.~\ref{fig1} that the IM transition becomes significant as the quantity 
$\hat{A}(0)/\vert\hat{A}(k_0)\vert$ is decreased.
For large values of $\hat{A}(0)/\vert\hat{A}(k_0)\vert$ (panels A and B), 
for a fixed value of the applied field, 
an increase of the temperature
leads to an increase of the symmetry of the equilibrium configuration.
However, a different phenomenology emerge in panels C, D, E and F, where the quantity $\hat{A}(0)/\vert\hat{A}(k_0)\vert$ 
takes 
values of $-0.6$ and $-0.8$.
In these cases, the low temperature regime of the phase diagrams is characterized by an IM phase boundary, in which 
a transition between 
the homogeneous and modulated phases is observed as the temperature is increased.

We quantify the extension of the IM by defining the following reentrance parameter: 
\begin{equation}
 R=\frac{h_\textrm{max}-h_0}{h_\textrm{max}},
 \label{R}
\end{equation}
where $h_0=h_c(0)$ is the critical field separating the modulated and homogeneous phase at zero temperature, while  
$h_\textrm{max}=h_c(T_\textrm{max})$ represent the maximum value of field along the phase boundary between these phases.
The parameter in Eq.~(\ref{R}) is zero for non-reentrant diagrams, and takes the value $1$ when $h_0=0$, when
the IM transition is observed for all fields of the phase boundary between homogeneous and modulated phases.

\begin{figure}[ht!]
    \centering
\includegraphics[width=1.0\columnwidth]{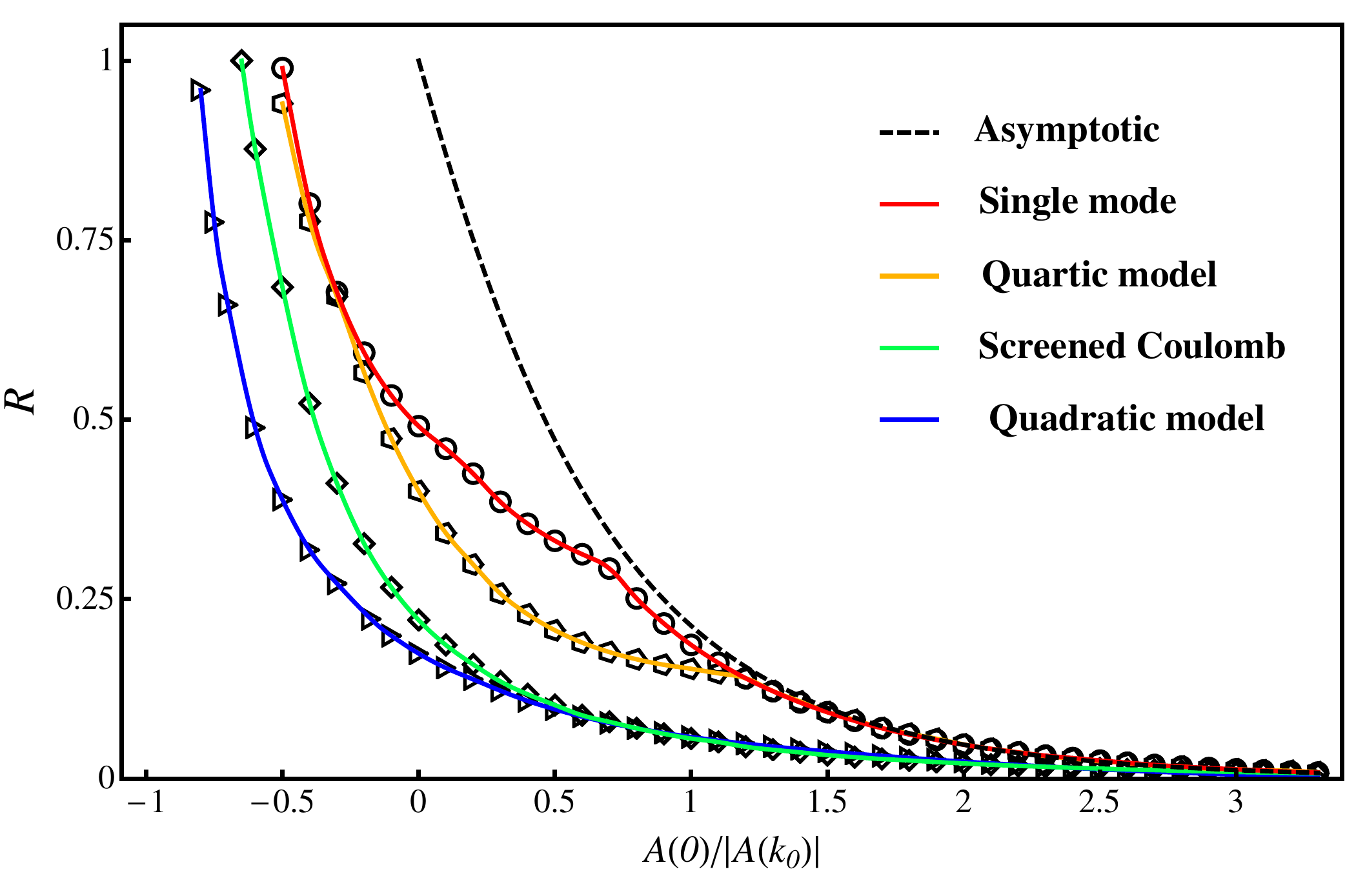}
\caption{Reentrant parameter as function of the relation $\hat{A}(0)/\vert\hat{A}(k_0)\vert$.
Solid curves correspond to the different interaction families presented in Table~\ref{table}.
The dashed line is the theoretical behavior valid in the asymptotic regime. 
}
\label{fig7}
\end{figure}

In Fig.~\ref{fig7} the reentrant parameter is shown for several families of interaction Hamiltonians.
As can be seen, the  previously observed behavior of the IM with respect to the relation  $\hat{A}(0)/\vert\hat{A}(k_0)\vert$ is confirmed for all 
types of interaction considered.
In all cases, this relation is able to tune the reentrant properties of the systems in the whole range $0\leq R\leq1$.
This result suggest that the IM is a universal feature, that can emerge in any particular type of interaction as far as the relation 
$\hat{A}(0)/\vert\hat{A}(k_0)\vert$ is sufficiently small.

It is also evident from Fig.~\ref{fig7} that, for a given value of the relation $\hat{A}(0)/\vert\hat{A}(k_0)\vert$, the single-mode approximation 
presents the highest reentrance among all the different families.
The reason of this finding relies in the fact that single-mode solutions implies a perfect sinusoidal structure. 
At small temperatures, the actual modulated solutions tend to be very structured, with important contribution of higher 
harmonics to lowering the free energy of the modulated solution.
Consequently, the single mode ansatz possesses the highest free energy among the 
different modulated solutions, and thus, the lower critical field separating the modulated and homogeneous phases. 

This difference of free energies/critical fields becomes less important as the temperature increases, since the 
weight of the higher harmonics decreases due to the effects of the entropic term in the free energy,
regardless the specific form of the interactions. 
Eventually, in the vicinity of $T_c$, the single mode solution is asymptotically exact.
In summary, among all solutions, the single mode one has the lowest $h_0$, while $h_{\mathrm{max}}$ is not particularly 
decreased. This explains why the single mode curve in Fig.~\ref{fig7} shows the highest reentrance for a given $\hat{A}(0)/\vert\hat{A}(k_0)\vert$, 
and is an upper bound for the different interaction families given the values $\hat{A}(0)$ and $\hat{A}(k_0)$.
In this way, the relation $\hat{A}(0)/\vert\hat{A}(k_0)\vert$ is the only needed input to know the maximum reentrance that any phase diagram can have.

The same arguments can be used to find a lower bound for the reentrance parameter of a given interaction $\hat{A}(k)$. 
Let us consider two fluctuation spectra $\hat{A}_1(k)$ and $\hat{A}_2(k)$, such that they have the same values 
of $\hat{A}(0)$ and $\hat{A}(k_0)$, 
and $\hat{A}_1(k)\leq\hat{A}_2(k)$ for all wave vectors. 
Then, $\hat{A}_1(k)$ will exhibit a lower free energy of the modulated solutions, due the contribution of higher harmonics. 
Again this difference is particularly important at lower temperatures, and decreases as temperature increases.
This implies that $\hat{A}_1(k)$ has higher critical fields, $h_c(T)$, separating the modulated and homogeneous phases, 
but this difference increases as temperature decreases. Consequently, the reentrance parameter for $\hat{A}_1(k)$ 
will be smaller then that for $\hat{A}_2(k)$. 
Among all the interaction families considered in table \ref{table}, 
the quadratic model has a fluctuation spectrum that limits inferiorly the other models\footnote{The quadratic model
stands beneath the screened Coulomb model only when $\hat{A}(0)\leq 0$} and 
as a consequence, it presents the lowest reentrance parameter in Fig.~\ref{fig1}. 

\subsection{Low Reentrance Regime}
As can be seen from Fig.~\ref{fig1}, for large values of $\hat{A}(0)/\vert\hat{A}(k_0)\vert$ the inverse transition appear just in a narrow region, corresponding to high external fields and low temperatures. 
In such conditions the transition between the modulated and the uniform phases is expected to occur with an average order parameter (magnetization) close to its saturation value. 
This implies that, close to the critical line, the modulations have small amplitude, which means that the transition should be 
continuous or, at most, weakly first order. 
Within this assumption an analytical study of the transition can be performed by using a single-mode description. 

According to Eqs.~(\ref{eq5}) and~(\ref{sol}), the transition between the modulated (bubble) and the uniform phase, 
in the high $\hat{A}(0)/\vert\hat{A}(k_0)\vert$ regime, must be governed by the effective free energy  
\begin{align}
H(c_0,c_1)&=\frac{1}{2}\hat{A}(0)c_0^2-\frac{3}{4}\vert\hat{A}(k_0)\vert c_1^2 \nonumber \\
 &+\frac{KT}{2}S(c_0)+KT\frac{3}{4}S^{(2)}(c_0)c_1^2 - h c_0,
 \label{smode}
\end{align}
where $c_0$ is the spatial average of the order parameter (e.g.~magnetization), and $c_1$ is the amplitude of 
the modulation. Additionally, $S^{(n)}(\phi)$ represent the derivative of order $n$ of the function $S(\phi)$. 
Performing the minimization process and enforcing the marginal stability of the modulated phase $\frac{\partial^2H}{\partial c_1^2}(c_0,c_1)=0$, we obtain
\begin{align}
c_{0,c}(T)&=S^{(2)}_{-1}\left(\frac{\vert\hat{A}(k_0)\vert}{KT}\right) \label{eq14a}\\
h_c(T)&=\hat{A}(0)c_{0,c}+KTS^{(1)}(c_{0,c}),
\label{eq14}
\end{align}
where $S^{(2)}_{-1}(\phi)$ represents the inverse function of $S^{(2)}(\phi)$. 
For our particular form of $S(\phi)$ the final expression of the critical line $h_c(T)$, separating the modulated from the uniform
phase, will be
\begin{align}
 h_c(T)&=\hat{A}(0)\sqrt{1-\frac{KT}{\vert\hat{A}(k_0)\vert}}\nonumber\\
 &+\frac{KT}{2}\log{\left(\frac{1+\sqrt{1-\frac{KT}{\vert\hat{A}(k_0)\vert}}}{1-\sqrt{1-\frac{KT}{\vert\hat{A}(k_0)\vert}}}\right)}.
 \label{eq15}
\end{align}

With the analytic form of the critical line it is possible then to obtain the behavior of the reentrant parameter in the asymptotic regime of large $\hat{A}(0)/\vert\hat{A}(k_0)\vert$.
According to its definition in Eq.~(\ref{R}) we have
\begin{equation}
 R=\frac{2\exp\left(-1-\frac{\hat{A}(0)}{\vert\hat{A}(k_0)\vert}\right)}{\frac{\hat{A}(0)}{\vert\hat{A}(k_0)\vert}
 +2\exp\left(-1-\frac{\hat{A}(0)}{\vert\hat{A}(k_0)\vert}\right)}.
 \label{Ra}
\end{equation}

For comparison, together with the results obtained by direct minimization of the mean-field free energy, the asymptotic 
behavior analytically obtained in Eq.~(\ref{Ra}) is presented in Fig.~\ref{fig7} with dashed lines.
As can be observed, the convergence to the asymptotic behavior occurs at relatively low values of $\hat{A}(0)/\vert\hat{A}(k_0)\vert$ for the quartic 
model and the single-mode interactions. 
These are precisely the fluctuation spectra with the highest energy cost for modes beyond the principal one,  
making these systems to form nearly single-mode profiles, which is the approximation assumed in Eq.~(\ref{smode}).

\section{Understanding the IM}
\label{understandIM}
\indent

A careful analysis of the results obtained in Eq.~(\ref{eq14}) and Eq.~(\ref{eq15}) confirm that, within the regime of 
large $\hat{A}(0)/\vert\hat{A}(k_0)\vert$, in order to produce a reentrant critical line, the entropic functional $S(\phi)$ must be steep enough in the 
vicinity of the saturation value of the order parameter. To understand this result it is important to notice the 
role of each term in Eq.~(\ref{eq14}). The first term represents an energetic contribution proportional
to the average value of the order parameter along the critical line, which is expected to increase towards saturation 
value as $T\to 0$ 
(see for example Eq.~(\ref{eq15})). This term accounts for a typical behavior of $h_c(T)$, and cannot produce a reentrant 
critical line.

On the other hand, the second term of Eq.~(\ref{eq14}), which represents the entropic 
contribution, is proportional to the first derivative of the entropic function with respect to the order parameter,
$S^{(1)}(c_{0,c})$,
evaluated at the average order parameter along the critical line.
To develop a reentrant $h_c(T)$ close to zero temperature, it is necessary that $S^{(1)}\left(c_{0,c}(T)\right)$ to be large 
enough as $c_{0,c}(T)$ approaches the saturation value in order to overcome the contribution of the first term in 
Eq.~(\ref{eq14}). 
In other words, the entropic function must be steep enough close to the saturation value of the order parameter in 
order to develop a reentrant behavior. It is worth recalling that even thought this result was obtained within certain 
approximations, still it suggests a central ingredient to observe IM in the considered models.

\begin{figure}[ht!]
    \centering
\includegraphics[width=1.0\columnwidth]{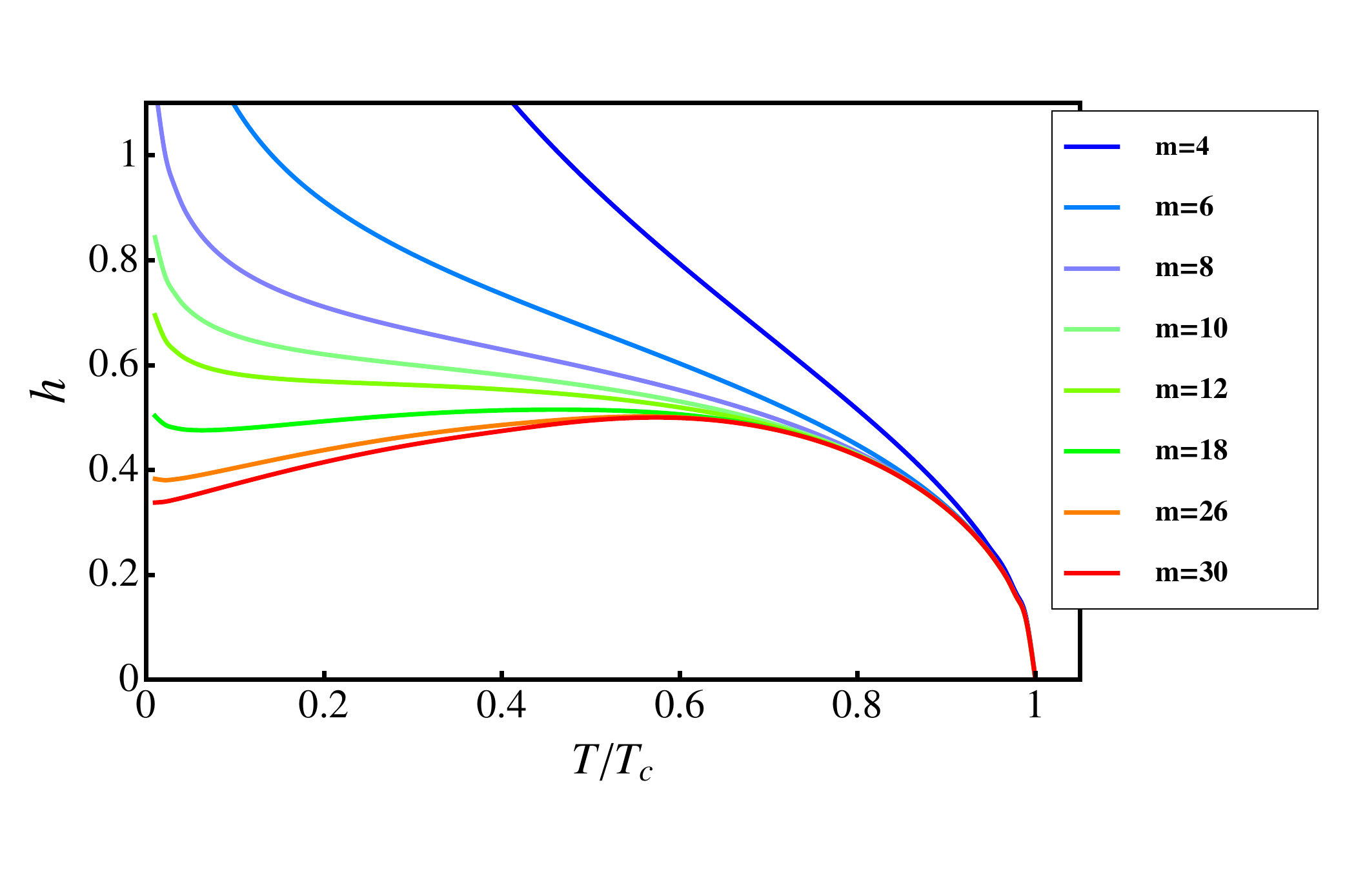}
\caption{Critical line $h_c(T)$ for a system in the single-mode approximation with $\hat{A}(0)=0$, where the local potential has been expanded in a Taylor series up to $m$ orders. 
Large values of $m$ are needed for the system to display IM features; this translates into a steeper entropic functional near the saturation value of $\phi$.
}
\label{fig8}
\end{figure}

In order to clarify the role of this ingredient, we calculate the critical line $h_c(T)$ for successive approximations 
to the entropic function, systematically increasing its steepness near the saturation value. 
We focus in the single-mode case, where the reentrance is the largest, and fix the value $\hat{A}(0)=0$.
We then expand the entropic function in a Taylor series up to order $m$ around $\phi=0$, where the larger the value of 
$m$, the steeper the entropic functional become around the saturation values $\phi=\pm 1$. The critical 
lines $h_c(T)$ for values $4\leq m\leq30$ are shown in Fig.~\ref{fig8}, where we can 
see that a reentrant behavior appears above $m\approx16$, and at $m\ge 30$ we can notice a reentrance similar to the 
ones observed experimentally.

Since the single-mode approximation is the case with highest reentrance for any $\hat{A}(0)/\vert\hat{A}(k_0)\vert$, the value $m=30$ represents 
the order of magnitude of the minimum power of the expansion of $S(\phi)$ that produces a reentrance in the whole 
low temperature regime. Even though this result is for the $\hat{A}(0)=0$ case, the same procedure could be extended 
to find a general result determining the minimal $m$ or steepness that produces the kind of reentrance 
in consideration, independently of the value $\hat{A}(0)$.
The above discussions shows that the IM phenomenology is dependent, at the same time, on the relative energy cost of the 
homogeneous phase ($\hat{A}(0)/\vert\hat{A}(k_0)\vert$, see Eq.~(\ref{Ra}) and Fig.~\ref{fig7}), and on the steep nature of the 
local potential around the saturation value of the order parameter.

In addition to the general ingredients discussed beforehand, one may wonder on the microscopic mechanism behind the 
reentrant behavior, or how the inverse melting actually takes place. 
For both homogeneous and modulated phases, along the critical line in relatively high fields, 
there is predominance of spatial regions where the order parameter is close to the saturation value. 
Let us consider a point in the phase diagram that belongs to the homogeneous phase, at the eminence of an IM  
transition to the modulated phase with a small increase of the temperature. 
Concerning the homogeneous solution, this increase in temperature will produce a decrease in its 
free energy, followed by a decrease of the average order parameter. 

In contrast, this increase in temperature produces a broadening of the interface between saturation values of the 
modulation - a natural process towards single mode solution at high temperatures. Because the entropy is a sharp 
function close to the saturation value, this small softening in the order parameter profile produces a large increase 
in entropy. It is important to stress that the regions that contributes most to this entropy change are those 
where the order parameter has experienced just a small change away from the saturation value. 
The latter is the mechanism behind the significant increase of the entropy of the modulated phase in relation 
to that of the homogeneous configuration, causing the IM transition. This comprehension is one
of the main results of this work.

Now is left to understand why reentrance diminishes monotonically as $\hat{A}(0)$ is increased, as can be observed in Fig.~\ref{fig7}. 
As expected when $\hat{A}(0)$ is increased there is an increasing energy cost of the homogeneous
phase. Consequently, for small
temperatures the critical fields separating the homogeneous and
modulated phases $h_c(T)$ become larger, and a higher external field
is needed to reach the homogeneous configuration. In the phase diagram
region in which the IM occurs, this implies a larger magnetization,
closer to the saturation value. Due to the interaction with the
external field, there is an increasing importance of the energetic
contribution to the free energy. As the role of the energy in the
low-temperature and high external field region of the phase diagram is
enhanced, the eventual free energy loss by the entropic mechanism in
the onset of the modulated state, increasing temperature, is
progressively less important. As a consequence, there is a monotonous
narrowing of the field range in which the IM occurs with the increase
of $\hat{A}(0)$, leading to the smaller values of $R$ observed in
Fig.~\ref{fig7}.

\section{Langevin Simulations}

In order to explore the extent of our theoretical results, it would be
interesting to compare them with simulations.  Indeed, these inverse
transitions in the presence of external fields have been explored
within microscopic models for dipolar frustrated ferromagnetic
materials, namely the dipolar Ising model \cite{dm10} and the dipolar
Heisenberg model with perpendicular anisotropy \cite{cannas11}.  On
the other hand, simulation of coarse-grained models \cite{NiSt2007,
  DiMeMuNiSt2011, MeStNi2015, NiMeSt2016} has also been used as an
alternative to study the long wavelength behavior of these models.

In general, all these numerical approaches have failed in reproducing the IM transition.
This is mainly due to the lack of a clear understanding of the features that a model has to include in order to develop that kind of behavior.
As discussed above, the effective model presented here in Eq.~(\ref{fef}) contains all the necessary ingredients to perform the reentrant 
transition with the appropriate set of parameters.
In the following, we will consider the quadratic model (see Table~\ref{table}) with several values of the curvature $a$.

The simulations are performed by numerical integration of the Langevin equation Eq.~(\ref{fef}). 
In contrast with the analytical treatment developed in the previous sections, the simulations consider all possible modes and fluctuations 
consistent with the system size.
In this way, the simulation also constitutes an ultimate proof of the validity of the analytical results obtained above.

The Langevin equation of motion for the effective model can be written in the form
\begin{eqnarray}
\nonumber
\frac{\partial\phi(\vec{x},t)}{\partial t} &=& T\  \mathrm{atanh}\{\phi(\vec{x},t)\} 
     +H-
     \big[A(k)\hat{\phi}(\vec{k},t)
     \big]_{\vec{x}}^{FT}
     \\
     && +\ \eta(\vec{x},t)
     \label{lan2}
\end{eqnarray}
where $]_{\vec{x}}^{FT}$ means the $\vec{x}$ component of the corresponding Fourier transform.
In order to properly deal with the stiffness of Eq.~(\ref{lan2}), due to the first term in the r.h.s., we implement the following
fully implicit first order scheme:
\begin{eqnarray}
\nonumber
\phi(\vec{x},t+dt) &=& \mathrm{tanh}
\Big\{
  -\frac{\phi(\vec{x},t+dt)}{T dt} 
  +\frac{1}{T} 
  \Big( \phi(\vec{x},t) 
  \\
     &+&H-
     \big[A(k)\hat{\phi}(\vec{k},t)
     \big]_{\vec{x}}^{FT}
     +\eta(\vec{x},t) 
     \Big)
     \Big\}
     \label{alg}
\end{eqnarray}
The above equation is discretized in a square lattice and solved on
each site by the Halley's method (\cite{PrTeVeFl1992}). Lattice
constant is chosen to be $dx=\pi/7$, so that the basic modulation
length spans 14 lattice sites. The linear size is $L=112$, such that
the system is able to accommodate 8 basic modulation lengths.

The three cases considered for the quadratic model have curvature
$a=0.1$, $0.4$ and $5.0$, corresponding to $\hat{A}(0)/\vert\hat{A}(k_0)\vert=-0.9$, $-0.6$ and
$4.0$, respectively (see Table~\ref{table}).  The time steps used in
Eq.~(\ref{alg}) for each case are $dt=0.1$, $0.05$ and $0.005$,
respectively. The estimated equilibration times are between $10^5$ and
$10^7$ time steps for the $a=0.1$ case, and between $10^4$ and $10^6$
time steps for the $a=0.4$ and $5.0$ cases. The phase diagrams were
constructed by slow cooling protocol at constant external fields.

\begin{figure}[ht!]
\includegraphics[width=0.85\columnwidth]{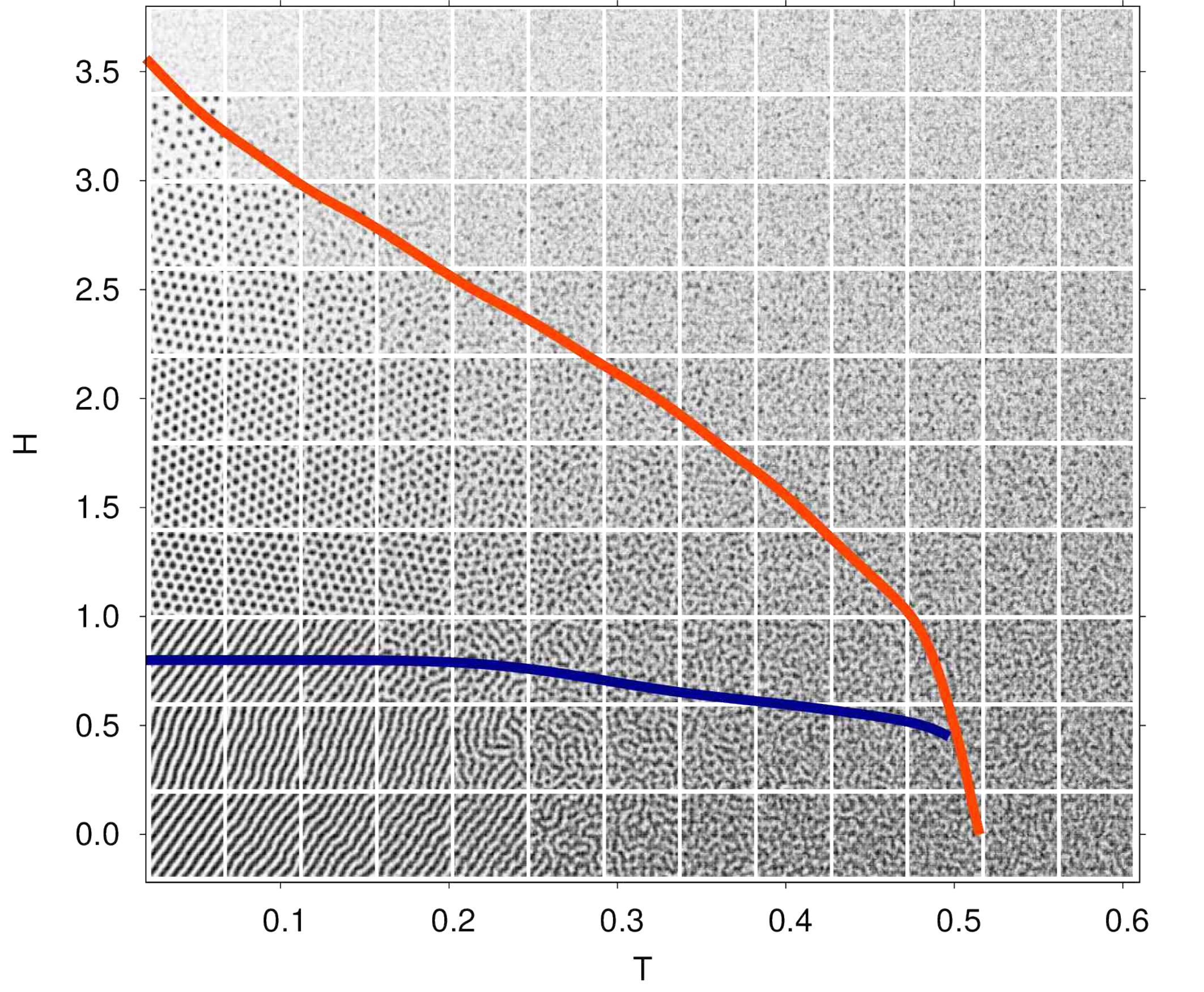}\\%
\includegraphics[width=0.85\columnwidth]{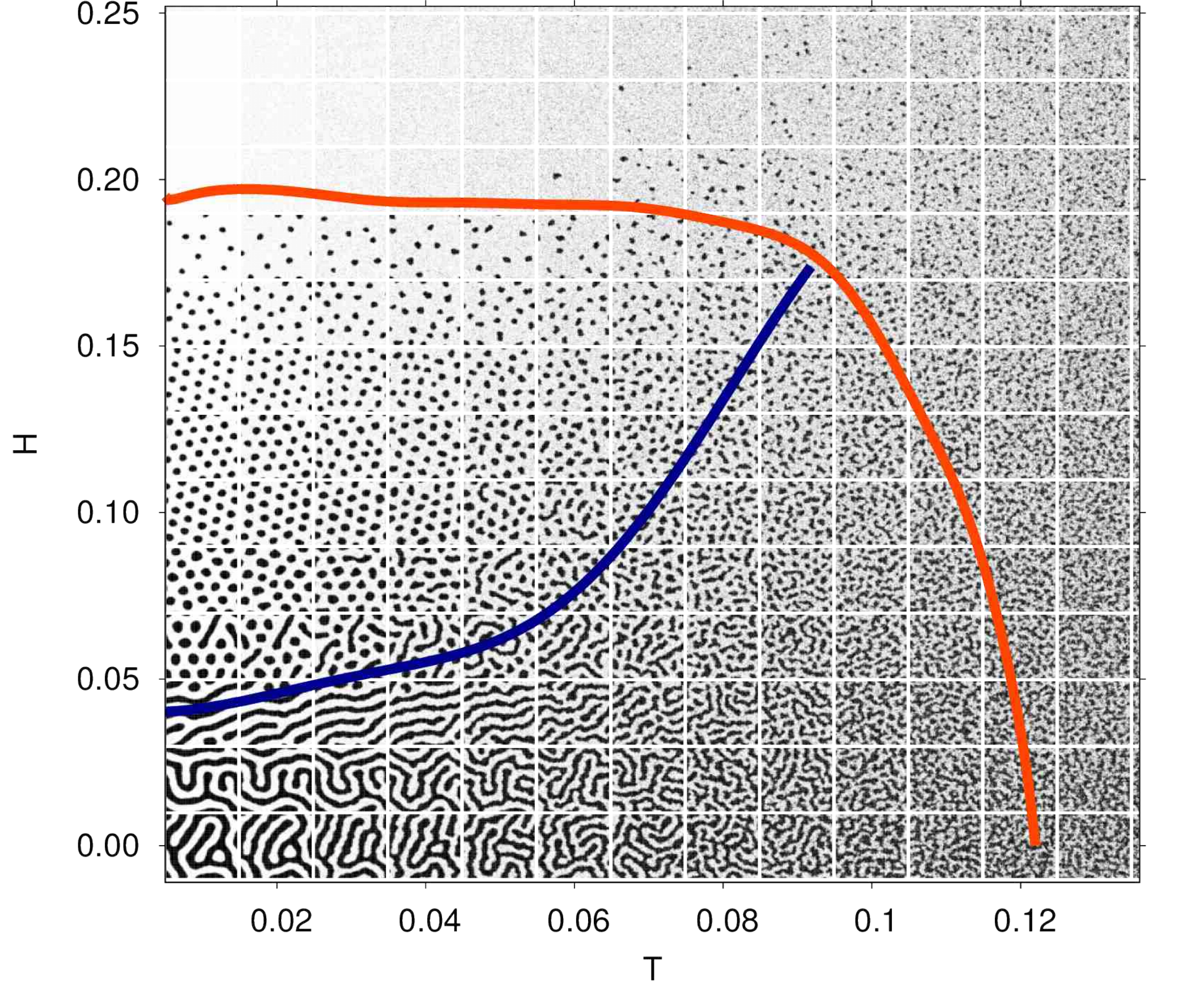}\\%
\includegraphics[width=0.85\columnwidth]{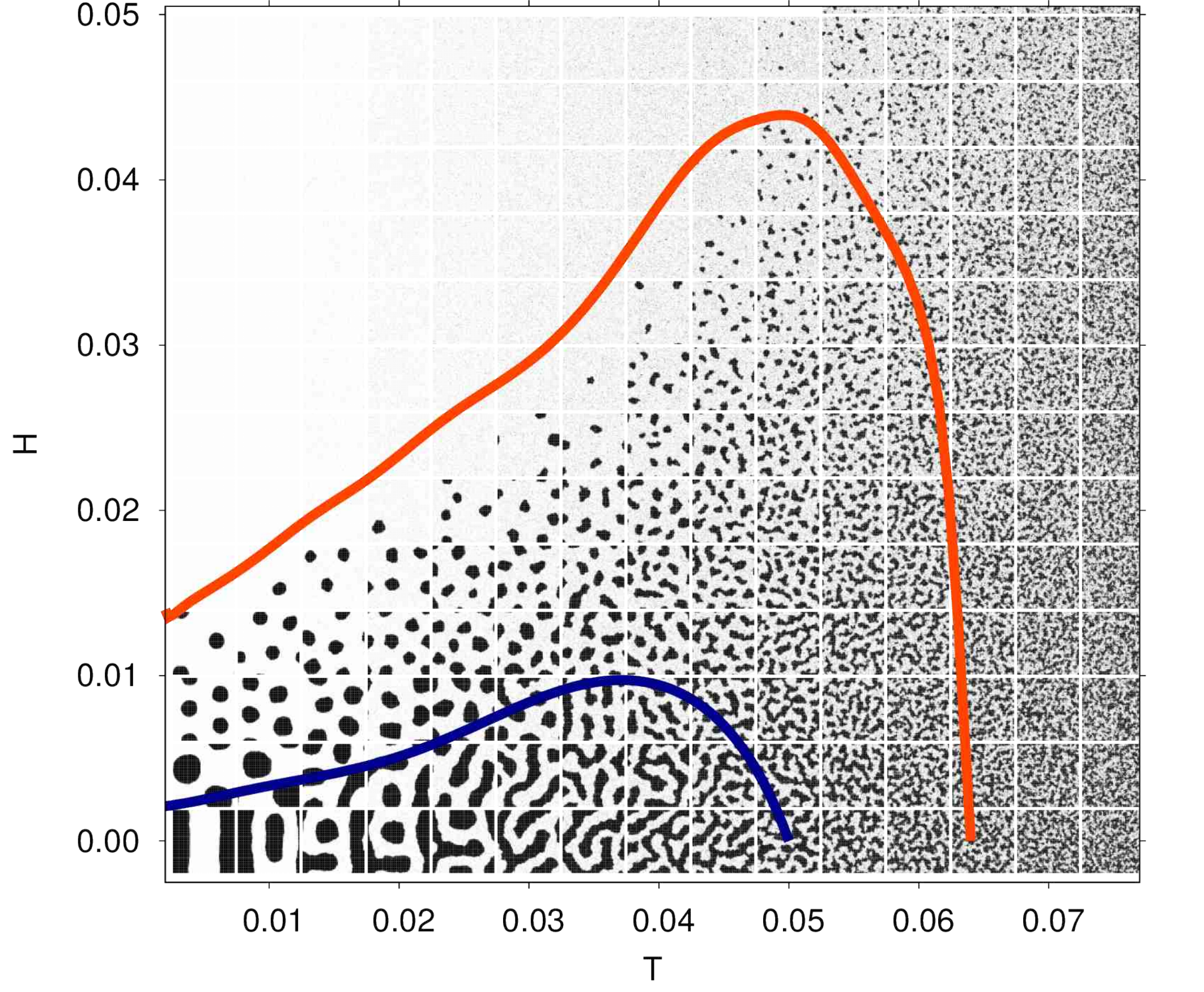}
\caption{Phase diagrams obtained by extensive Langevin simulations of
  a system interacting via a quadratic fluctuation spectrum. The upper
  panel correspond to $\hat{A}(0)/\vert\hat{A}(k_0)\vert=4.0$, followed by $\hat{A}(0)/\vert\hat{A}(k_0)\vert=-0.6$ and
  $\hat{A}(0)/\vert\hat{A}(k_0)\vert=-0.9$. Configurations indicate the representative state for
  each $(T,H)$ point indicated by the center of the configuration.
  The actual resolution of points explored in the simulations is much
  finer, and was used to estimate the modulated/non-modulated (red)
  and stripe/bubble (blue) separation lines (see text).
Notice that both these criteria does not distinguish the degree of
order, so the lines do not correspond necessarily to phase
transitions.}
\label{sim}
\end{figure}

In order to identify the region of where modulation sets in, we have
estimated the $\left<\phi(0)\phi(\vec{x})\right>$ correlation length
through a nonlinear fit of the circularly averaged structure factor
to $\hat{S}^{-1}(k) = b(k-k_0)^2+r$, so that the correlation length is
given by $\xi=\sqrt{b/r}$. Thus, the modulated region of the phase
diagram is defined here as the region in which the correlation length
of the system is larger than half of the basic modulation length, i.e.
$\xi>7$. This estimate of the crossover between modulated and
homogeneous/non-modulated regions is depicted in Fig.~\ref{sim} with a
red line.

In addition to the modulated region, it is interesting to
differentiate regions dominated by bubbles or stripes. This can be
done quantitatively by measuring the area $A$ and perimeter $L$
defined by the $\phi(\vec{x})=\left<\phi\right>$ closed
contours. Controlling the averaged value of the quantity $4\pi A/L^2$
(which is 1 for a circle, and formally 0 for an infinite stripe), if
its resulting value is greater than $\frac12$, then we consider that
the configurations are dominated by bubbles. Otherwise, they are
dominated by stripes. This estimate of the crossover between stripe
dominated and bubble dominated modulation patterns is depicted in each
case in Fig.~\ref{sim} with a blue line.

In Fig.~\ref{sim} the phase diagrams obtained by the Langevin
simulations are presented for several values of the ratio $\hat{A}(0)/\vert\hat{A}(k_0)\vert$.
As can be seen, the IM is encountered in the simulation and its behavior with respect to the specific value of $\hat{A}(0)$ 
is virtually the same as that predicted with the analytical approximations in Fig.~\ref{fig1}.
That is, smaller values of $\hat{A}(0)/\vert\hat{A}(k_0)\vert$ imply larger reentrance.
These results are a further support to the validity of the analytical outcomes.

For small temperatures the values of the critical fields observed in simulations and mean-field approach are very similar. 
For higher temperatures, however, the increasing role of the fluctuations makes the critical fields obtained in the simulations
to remain below its mean-field counterparts.
Consequently, the extent of the reentrance observed in the simulations is smaller than that observed within the mean-field approximations.
Furthermore, thermal fluctuations strongly decrease the temperature window in which the modulated phases appear in comparison with the one expected in 
the mean field approximations. 
It is important to stress out that while in the mean-field diagrams
the lines corresponds to phase transitions, here it is not necessarily
the case. For example, there is no symmetry breaking taking place
between a disordered modulated phase and a homogeneous phase.

\section{Conclusions}
\label{conc}

\mbox{}
\indent

In this report we have addressed the problem of the IM transition in
the frame of a coarse-grained model with generic interactions.  A
detailed characterization of the different phase diagrams was obtained
by means of a minimization-variational technique and the use of the
mean-field form for the local entropic contribution.  We identified
two fundamental ingredients for the IM to take place. First, it is
necessary a low enough energy cost of the homogeneous phase,
relatively to the modulated phase. This is achieved whenever the
non-local repulsive interaction is much weaker than the attractive
local interaction. The second ingredient is that the local order
parameter must be limited. This can be achieved either naturally, in
systems where the microscopic variables are intrinsically limited,
like spins in a magnetic material, or effectively, like in a fluid
where density can be limited due to the presence of a hard-core
potential.

Furthermore, the microscopic mechanism behind the reentrant behavior
is closely related to these ingredients. For fixed external fields, as
we increase temperature, the modulated solution undergoes a natural
processes of softening that diminishes the regions in which the order
parameter is close to the saturation value. This in turn can produce a
significant entropy gain if the local entropy function is steep enough
close to the saturation value. Whenever the free energy loss
associated to this mechanism is high enough, the system undergoes a IM
transition from the homogeneous to the modulated phase, as we increase
temperature.

It is worth mentioning that our results are in agreement
with previous works, in particular our conclusion that the reentrance
parameter decays exponentially with the ratio
$\hat{A}(0)/\vert\hat{A}(k_0)\vert$ explain why, when repulsive
interactions of the form $r^{-\alpha}$ are present, the IM does not
occurs if $\alpha$ is lower than the system's dimension
\cite{Po2010}.  In this case it is simple
to show that $\hat{A}(0)=+\infty$, which immediately justify the
previous conclusion.  Regarding the mechanism behind the IM, some
authors \cite{velasque14,mc2016} have
pointed out that when IM takes place, there is a significant increase
of the domain wall width, as temperature is raised at constant
field. This variation of the domain wall width is reflected in an
entropy gain which ultimately justify the reentrance to a modulated
phase. In the present work we have generalized these previous results
by understanding that the entropy gain takes place mainly due the
shrink of the saturated regions in the modulation profile, as
temperature increases. This explain the appearance of IM not only in
the previously studied cases but also in the single mode phase
diagram, where IM takes place without changing the domain wall width.

The robustness of our findings were tested by considering different
models, all of them showing excellent agreement with our general
predictions. In the limit of large $\hat{A}(0)/\vert\hat{A}(k_0)\vert$, the different models
presents a similarly small extension of the IM in the phase diagram,
showing good agreement with our analytical approximation.
For arbitrary isotropic competing interactions, we have also found a
lower and an upper bound to the extension of the IM in the phase
diagram. For fixed values of $k_0$, $\hat{A}(k_0)$ and $\hat{A}(0)$,
the reentrance associated to an arbitrary fluctuation spectrum will be
smaller than the corresponding in the single-mode approximation, and
larger than that of a quadratic model, whenever this is a lower
envelope for the fluctuation spectrum.

As a further validation of the minimization-variational technique, a
numerical scheme was developed for the Langevin equation of this
system and the simulation results qualitatively confirm the main
claims regarding the shape of the phase diagrams. The numerical phase
diagrams shows a number of different phases within the modulated
regions that are beyond the scope of the mean-field
approximations. 
The nature of these phases and the construction of a
more detailed phase diagram will be addressed in a future work.

\section*{Acknowledgement}
We acknowledge resources provided by the Swedish National Infrastructure for Computing (SNIC) at HPC2N.
RDM acknowledges travel grant from Roland Gustafssons Stiftelse f\"or teoretisk fysik.



\bibliographystyle{unsrt}

\end{document}